\documentclass[11pt]{article}
\usepackage[totalwidth=460truept,totalheight=600truept]{geometry}
\usepackage{epsfig,latexsym,amssymb}

\def\theequation{\arabic{section}.\arabic{equation}}

\renewcommand{\theequation}{\thesection.\arabic{equation}}
\global\arraycolsep=1truept

\begin{document}

\hfill \hfill IFUP-TH 2007-8

\vskip 1.4truecm

\begin{center}
{\huge \textbf{Improved Schwinger-DeWitt Techniques}}

{\large \textbf{\vskip .1truecm}}

{\huge \textbf{For Higher-Derivative Perturbations }}

{\large \textbf{\vskip .1truecm}}

{\huge \textbf{Of Operator Determinants}}

\vskip 1.5truecm

\textsl{{Damiano Anselmi} }

\textsl{\textit{{Dipartimento di Fisica ``Enrico Fermi'', Universit\`{a} di
Pisa, } }}

\textsl{\textit{{Largo Bruno Pontecorvo 3, I-56127 Pisa, Italy, } }}

\textsl{\textit{{and INFN, Sezione di Pisa, Pisa, Italy} }}

\textsl{\textit{damiano.anselmi@df.unipi.it }}

\textsl{\textit{\bigskip }}

\textsl{\textit{and }}

\textsl{\textit{\bigskip }}

\textsl{\textit{{Anna Benini} }}

\textsl{\textit{{Department of Mathematics, University of Toronto} }}

\textsl{\textit{{40 St. George Street, Toronto, Ontario, Canada M5S 2E4} }}

\textsl{\textit{anna@math.utoronto.ca }}
\end{center}

\vskip 1truecm

\begin{center}
\textbf{Abstract}
\end{center}

\bigskip

{\small We consider higher-derivative perturbations of quantum gravity and
quantum field theories in curved space and investigate tools to calculate
counterterms and short-distance expansions of Feynman diagrams. In the case
of single higher-derivative insertions we derive a closed formula that
relates the perturbed one-loop counterterms to the unperturbed
Schwinger-DeWitt coefficients. In the more general case, we classify the
contributions to the short-distance expansion and outline a number of
simplification methods. Certain difficulties of the common differential
technique in the presence of higher-derivative perturbations are avoided by
a systematic use of the Campbell-Baker-Hausdorff formula, which in some
cases reduces the computational effort considerably. }

\vfill\eject

\section{Introduction}

\setcounter{equation}{0}

In quantum gravity infinitely many independent couplings are necessary to
remove the divergences. Practical tools to do systematic computations with
non-polynomial lagrangians are not available today. In this paper we
investigate techniques to express the one-loop counterterms of the most
general lagrangian in closed form. Consider a generic gravitational action
constructed with the curvature tensors and their derivatives, 
\[
S_{\mathrm{tree}}=\int \mathcal{L}_{\mathrm{tree}}. 
\]
The one-loop counterterms are collected in a functional $S_{\mathrm{1 div}}$
uniquely determined by $S_{\mathrm{tree}}$, 
\[
S_{\mathrm{1 div}}=\int \mathcal{L}_{\mathrm{1 div}}=S_{\mathrm{1 div}}(S_{%
\mathrm{tree}}). 
\]
If $S_{\mathrm{1 div}}(S_{\mathrm{tree}})$ could be written explicitly, it
would be possible to search for special $S_{\mathrm{tree}}$'s containing a
finite number of independent couplings, such that 
\[
S_{\mathrm{tree}}-S_{\mathrm{1 div}}(S_{\mathrm{tree}})=S_{\mathrm{tree}%
}^{\prime }, 
\]
up to two-loop corrections, where $S_{\mathrm{tree}}^{\prime }$ coincides
with $S_{\mathrm{tree}}$ up to redefinitions of fields and couplings. If $S_{%
\mathrm{tree}}$ were so special to satisfy analogous identities for the
two-loop and higher-order counterterms, then it would define a
``renormalizable'' theory.

The search for renormalizable theories beyond power counting is not an easy
task, but can teach us a lot about the structure of counterterms and their
classification. In quantum gravity renormalization turns on vertices with
dimensionalities greater than four. It is well-known that in the absence of
matter, the one-loop counterterms can be eliminated with a field
redefinition of the metric tensor \cite{thooftveltman}. The first new vertex
is cubic in the Riemann tensor and removes a two-loop divergence \cite
{sagnotti,vdv}. The corrected quantum gravity lagrangian reads 
\begin{equation}
S_{\mathrm{QG}}=\frac{1}{2\kappa ^{2}}\int \sqrt{-g}R+\lambda \int \sqrt{-g}%
R_{\mu \nu \rho \sigma }R^{\mu \nu \alpha \beta }R_{\alpha \beta }^{\rho
\sigma }+\mathcal{O}(R^{4}).  \label{corra}
\end{equation}
Expanding (\ref{corra}) around a background metric, the one-loop Feynman
diagrams are encoded in the determinant of a differential operator
containing higher-derivative terms. In general, the higher-derivative terms
can be treated perturbatively or non-perturbatively. In the former approach 
\cite{thooftveltman,sagnotti,vdv} (``quantum gravity'') they are viewed as
perturbations of the Einstein lagrangian: the theory is non-renormalizable,
but perturbatively unitary. In the latter approach \cite{stelle,fradkin,pron}
(``higher-derivative gravity'') they are used to improve the behavior of
Green functions at short distances: the theory is renormalizable, but not
unitary. Here we are interested in the former approach, which is equivalent
to study the insertions of higher-derivative operators in the Feynman
diagrams of quantum gravity. Observe that in (\ref{corra}) higher powers of
the curvature tensor generate perturbations with an arbitrary number of
derivatives.

We illustrate our techniques in the case of a scalar operator of the form 
\begin{equation}
\widehat{H}=\widehat{H}_{0}+\widehat{H}_{1},\qquad \widehat{H}_{0}=\Box -\xi
R,\qquad \widehat{H}_{1}=\sum_{n=0}^{\infty }V^{\mu _{1}\cdots \mu
_{n}}D_{\mu _{1}}\cdots D_{\mu _{n}},  \label{h1}
\end{equation}
For our purposes $\widehat{H}_{1}$ can be treated perturbatively. We loose
no generality if we assume that the tensors $V^{\mu _{1}\cdots \mu _{n}}$
are completely symmetric. Indeed, commuting the covariant derivatives every
antisymmetric component of $V^{\mu _{1}\cdots \mu _{n}}$ can be reduced to a
combination of $V$-terms with fewer indices. We investigate tools to study
the perturbative expansion of the $\widehat{H}$-determinant and simplify the
computation of its coefficients. Our arguments are general and their
extension to spinors, spin-1 fields and the graviton is direct. The case of
gravity is addressed.

Calculations in quantum gravity are conveniently done using the background
field method \cite{back,ab} and, at the one-loop level, the Schwinger-DeWitt
techniques \cite{schwinger,dewitt}, because they manifestly preserve
covariance. The common approach is to perform a Schwinger-DeWitt expansion
of the Green function, derive a differential equation for its coefficients
and work out their short-distance expansion by repeated differentiation.
However, in the presence of higher-derivative perturbations the differential
approach has some difficulties, which can be easily overcome working at the
level of operators and systematically using the Campbell-Baker-Hausdorff
(CBH) formula before taking the coincidence limits. In some cases this
approach simplifies the calculation considerably and allows the derivation
of some closed formulas. Moreover, it singles out that a number of involved
expressions are just total derivatives and so can be neglected for the
purposes of renormalization.

The paper is organized as follows. In section 2 we recall the
Schwinger-DeWitt approach and explain the difficulties of the differential
technique. In section 3 we introduce the CBH approach and derive the closed
counterterm formulas (\ref{genfor}) and (\ref{notrace}) for the most general
single insertions. We also prove the new identity (\ref{p1}). In section 4
we describe the method in general and make explicit computations. In
particular, formula (\ref{m3}) for three-derivative perturbations is a new
result. In section 5 we describe how the techniques apply to gravity.
Section 6 contains our conclusions.

\section{Difficulties of the differential approach}

\setcounter{equation}{0}

Given an operator $\widehat{H}$, define the function $H(s;x,x^{\prime })$, $%
s>0$, as the solution of the equation 
\begin{equation}
\left( i\frac{\partial }{\partial s}+\widehat{H}\right) H(s;x,x^{\prime })=0
\label{diffa}
\end{equation}
with the boundary condition 
\begin{equation}
H(0;x,x^{\prime })=\frac{1}{\sqrt{-g(x)}}\delta ^{(4)}(x-x^{\prime }).
\label{conda}
\end{equation}
If $\widehat{H}$ has for example the form 
\begin{equation}
\widehat{H}_{0}=\Box -\xi R,  \label{h0}
\end{equation}
where $\Box $ denotes the covariant D'Alembertian, the Schwinger-DeWitt
expansion of the associated function $H_{0}(s;x,x^{\prime };\xi )$ reads 
\begin{equation}
H_{0}(s;x,x^{\prime };\xi )=-\frac{i}{(4\pi )^{2}s^{2}}\mathrm{\exp }\left( 
\frac{i\sigma (x,x^{\prime })}{2s}\right) \sum_{n=0}^{\infty
}(is)^{n}A_{n}(x,x^{\prime };\xi ),  \label{expa}
\end{equation}
with the boundary condition 
\begin{equation}
\lim_{x^{\prime }\rightarrow x}A_{0}(x,x^{\prime };\xi )=1.  \label{bounda}
\end{equation}
In (\ref{expa}) $\sigma (x,x^{\prime })$ is one half the squared geodesic
distance between $x$ and $x^{\prime }$ and satisfies 
\begin{equation}
\frac{1}{2}\sigma ^{;\mu }\sigma _{;\mu }=\sigma ,\qquad \sigma (x,x)=\sigma
_{;\mu }(x,x)=0,\qquad \sigma _{;\mu \nu }(x,x)=g_{\mu \nu }(x).
\label{sigma}
\end{equation}
Equation (\ref{diffa}), with $H\rightarrow H_{0}$, generates a differential
recursion relation for the coefficients $A_{n}$, $n\geq 0$, namely 
\begin{equation}
(n-2)A_{n}+\sigma ^{;\mu }A_{n;\mu }+\frac{1}{2}A_{n}\Box _{x}\sigma =\left(
\Box _{x}-\xi R\right) A_{n-1},  \label{diffe}
\end{equation}
with $A_{-1}=0$. By repeated differentiation, the recursion relation (\ref
{diffe}) can be used to calculate the short-distance expansion of the
Schwinger-DeWitt coefficients $A_{n}(x,x^{\prime };\xi )$. For this purpose,
it is sufficient to compute the coincidence limits $A_{n}(x,x;\xi )$, which
are called ``diagonal coefficients'', and the coincidence limits of the
covariant derivatives of $A_{n}(x,x^{\prime };\xi )$, which are called
``off-diagonal coefficients''. The calculational method just described will
be called the ``DeWitt differential approach''. The coincidence limits will
be denoted with an overline. The first two coefficients, $\overline{A_{1}}$
and $\overline{A_{2}}$ have been computed by DeWitt in \cite{dewitt}, $%
\overline{A_{3}}$ by Sakai in \cite{sakai} and Gilkey in \cite{gilkey}, $%
\overline{A_{4}}$ by Amsterdamski, Berkin and O'Connors in \cite{amster} and
Avramidi in \cite{avramidi}, $\overline{A_{5}}$ by van de Ven in \cite
{vandeven}. In \cite{decanini} the first off-diagonal coefficients have been
recently worked out to a considerable order and number of derivatives.

The one-loop contributions to the generating functional $\Gamma $ of
one-particle irreducible functions read 
\begin{equation}
\Gamma ^{(1)}=-\frac{i}{2}\int_{\delta }^{\infty }\frac{\mathrm{d}s}{s}\int 
\mathrm{d}^{4}x\sqrt{-g(x)}H(s;x,x;\xi ).  \label{gam}
\end{equation}
Although our techniques are general, we focus on the scheme-independent
(logarithmic) divergences. In the notation commonly used in dimensional
regularization ($\ln \delta \rightarrow -1/(2\varepsilon )$), we have 
\begin{equation}
\Gamma _{\mathrm{div}}^{(1)}=\frac{1}{64\pi ^{2}\varepsilon }\int \sqrt{-g}\ 
\overline{A}_{2}.  \label{div}
\end{equation}

Thus, to study the one-loop renormalization one has to calculate the
coincidence limit of the second Schwinger-DeWitt coefficient. In the DeWitt
differential approach, this goal can be achieved repeatedly differentiating
equation (\ref{diffe}) and the first of (\ref{sigma}), and taking
coincidence limits with the help of (\ref{bounda}) and (\ref{sigma}).

However, the differential approach is not convenient to study
higher-derivative perturbations. The reason can be appreciated already in
flat space. Consider 
\begin{equation}
\widehat{H}=\widehat{H}_{0}+\widehat{H}_{1},\qquad \widehat{H}_{0}=\partial
^{2},\qquad \widehat{H}_{1}=\lambda (\partial ^{2})^{2}.  \label{flat}
\end{equation}
The unperturbed flat-space Green function reads 
\[
H_{0}(s;x-x^{\prime })=-\frac{i}{(4\pi )^{2}s^{2}}\mathrm{\exp }\left( \frac{%
i(x-x^{\prime })^{2}}{4s}\right) . 
\]
The first observation is that the Schwinger-DeWitt expansion (\ref{expa}) of 
$H(s;x-x^{\prime })$ needs to be replaced with a sum containing arbitrary
negative powers of $s$, namely 
\begin{equation}
H(s;x)=-\frac{i}{(4\pi )^{2}s^{2}}\mathrm{\exp }\left( \frac{ix^{2}}{4s}%
\right) \sum_{n=-\infty }^{\infty }(is)^{n}B_{n}(\lambda ,x).  \label{ezpa}
\end{equation}
Nevertheless, at each order in $\lambda $ the sum is bounded from below. In
particular, at $\mathcal{O}(\lambda )$ the sum starts at $n=-3$. To this
order, equation (\ref{diffa}) gives the relations 
\[
-\frac{\lambda (x^{2})^{2}}{16}=3B_{-3}-x^{\mu }\partial _{\mu }B_{-3},\quad 
\frac{3\lambda }{2}x^{2}=2B_{-2}+\Box B_{-3}-x^{\mu }\partial _{\mu
}B_{-2},\quad -6\lambda =B_{-1}+\Box B_{-2}-x^{\mu }\partial _{\mu }B_{-1}. 
\]
Differentiating these relations a suitable number of times and taking the
coincidence limits ($x\rightarrow 0$), we find $\overline{B_{-2}}=\overline{%
B_{-3}}=\overline{\Box B_{-3}}=0$, plus the relations 
\begin{equation}
\overline{\Box ^{2}B_{-3}}=12\lambda ,\qquad \overline{B_{-1}}+\overline{%
\Box B_{-2}}=-6\lambda ,  \label{rela}
\end{equation}
which are valid up to higher orders in $\lambda $. Two equations give the
first of (\ref{rela}), so one quantity, $\overline{\Box B_{-2}}$, remains
undetermined.

More generally, the recurrence relations for the coefficients $B_{-k}$ with $%
k>0$ have the form 
\begin{equation}
kB_{-k}-x^{\mu }\partial _{\mu }B_{-k}=P_{k}+\mathcal{O}(\lambda ),
\label{recurso}
\end{equation}
where $P_{k}$ possibily depends on $\partial^{k^{\prime}-k}B_{-k^{\prime }}$
with $k^{\prime }>k$. The left-hand side of (\ref{recurso}) vanishes, in the
coincidence limit, when $k$ derivatives act on it. Therefore (\ref{recurso})
does not provide information about $\overline{\partial ^{k}B_{-k}}$. This
ambiguity has the following explanation. The initial condition (\ref{conda})
determines the solution uniquely. While in the unperturbed problem (\ref
{conda}) is exhaustively expressed by (\ref{bounda}), in the perturbed
problem it is expressed by $\overline{B}_{0}=1$ plus suitable relations
among the $B_{n}$'s with $n<0$. The extra relations, however, are not
immediately readable from the expansion (\ref{ezpa}), because of the
negative powers of $s$ contained in the sum, and need to be worked out
independently.

For these reasons it is more convenient to pursue a strategy that
incorporates the boundary condition (\ref{conda}) automatically. This goal
is achieved writing 
\begin{equation}
H(s;x,x^{\prime })=\langle x\mid \mathrm{e}^{i\widehat{H}s}\mid x^{\prime
}\rangle ,  \label{acca}
\end{equation}
where $\mid x\rangle $ are position eigenstates, $\widehat{x}^{\mu }\mid
x\rangle =x^{\mu }\mid x\rangle $, $\langle x^{\prime }\mid x\rangle =\delta
(x^{\prime }-x)$. Noting that in the case (\ref{flat}) $\widehat{H}_{0}$ and 
$\widehat{H}_{1}$ commute, we can write 
\[
H(s;x,x^{\prime })=\langle x\mid \mathrm{e}^{i\widehat{H}_{0}s}\mathrm{e}^{i%
\widehat{H}_{1}s}\mid x^{\prime }\rangle =\mathrm{e}^{i\widehat{H}%
_{1}s}H_{0}(s;x,x^{\prime })=-\frac{i}{(4\pi )^{2}s^{2}}\mathrm{e}%
^{is\lambda (\partial ^{2})^{2}}\mathrm{\exp }\left( \frac{i(x-x^{\prime
})^{2}}{4s}\right) . 
\]
This procedure does not contain any ambiguity, and easily leads to 
\[
\overline{B_{-1}}=6\lambda ,\qquad \overline{\Box B_{-2}}=-12\lambda . 
\]

In the rest of the paper we use this strategy in curved space. We name it
``CBH approach'', because it involves a systematic use of the CBH\ formula.
Besides avoiding the difficulty just mentioned, in some cases the CBH
approach reduces the calculational effort considerably. Moreover, it allows
us to calculate each coefficient $B_{n}$ directly, without having first to
recursively calculate the $B_{m}$'s with $m<n$.

\section{The CBH approach}

\setcounter{equation}{0}

In curved space, formulas (\ref{acca}) and (\ref{ezpa}) are replaced by 
\begin{eqnarray}
H(s;x,x^{\prime };\xi ) &=&(-g(x))^{-1/4}\langle x\mid \mathrm{e}^{i%
\widetilde{H}s}\mid x^{\prime }\rangle (-g(x^{\prime }))^{-1/4}  \nonumber \\
&=&-\frac{i}{(4\pi )^{2}s^{2}}\mathrm{\exp }\left( \frac{i\sigma
(x,x^{\prime })}{2s}\right) \sum_{n=-\infty }^{\infty
}(is)^{n}B_{n}(x,x^{\prime };\xi ),  \label{sdw}
\end{eqnarray}
where $\widetilde{H}=(-g)^{1/4}\widehat{H}(-g)^{-1/4}$. Define $\widehat{H}%
_{0}$ and $\widehat{H}_{1}$ as in (\ref{h1}) and write $\widetilde{H}=%
\widetilde{H}_{0}+\widetilde{H}_{1}$. The CBH formula reads 
\begin{equation}
\mathrm{e}^{i\widetilde{H}s}=\mathrm{e}^{i\widetilde{H}_{0}s}\sum_{n=0}^{%
\infty }\frac{(is)^{n}}{n!}\int_{0}^{1}\mathrm{d}\zeta _{1}\cdots \mathrm{d}%
\zeta _{n}\ \mathrm{T}\left[ \widetilde{H}_{1}(\zeta _{1})\cdots \widetilde{H%
}_{n}(\zeta _{n})\right] ,  \label{CBH}
\end{equation}
where T denotes the ordered product and 
\begin{equation}
\widetilde{H}_{1}(\zeta )=\mathrm{e}^{-i\widetilde{H}_{0}s\zeta }\widetilde{H%
}_{1}\mathrm{e}^{i\widetilde{H}_{0}s\zeta }=\sum_{n=0}^{\infty }\frac{%
(-is\zeta )^{n}}{n!}(\mathrm{ad}\widetilde{H}_{0})^{n}\widetilde{H}_{1}.
\label{sum}
\end{equation}
with $(\mathrm{ad}A)B\equiv [A,B]$. Consider, for example, the first order
in $\widetilde{H}_{1}$, namely the diagrams that contain a single insertion
of the perturbation. We have 
\begin{eqnarray*}
H(s;x,x^{\prime };\xi ) &=&H_{0}(s;x,x^{\prime };\xi )+is\int_{0}^{1}\mathrm{%
d}\zeta (-g(x))^{-1/4}\langle x\mid \mathrm{e}^{i\widetilde{H}_{0}s(1-\zeta
)}\widetilde{H}_{1}\mathrm{e}^{i\widetilde{H}_{0}s\zeta }\mid x^{\prime
}\rangle (-g(x^{\prime }))^{-1/4} \\
&&+\mathcal{O}(H_{1}^{2}).
\end{eqnarray*}
The one-loop contributions (\ref{gam}) to the $\Gamma $ functional become 
\[
\Gamma ^{(1)}=\frac{1}{2}\int_{\delta }^{\infty }\mathrm{d}s\int_{0}^{1}%
\mathrm{d}\zeta \int \mathrm{d}^{4}x\ \langle x\mid \mathrm{e}^{i\widetilde{H%
}_{0}s(1-\zeta )}\widetilde{H}_{1}\mathrm{e}^{i\widetilde{H}_{0}s\zeta }\mid
x\rangle .
\]
The $\zeta $-integrand is just the trace of the operator contained between
the bra and the ket. We can use the ciclicity of the trace and get 
\begin{equation}
\Gamma ^{(1)}=\frac{1}{2}\int_{\delta }^{\infty }\mathrm{d}s\int \mathrm{d}%
^{4}x\ \langle x\mid \widetilde{H}_{1}\mathrm{e}^{i\widetilde{H}_{0}s}\mid
x\rangle =\frac{1}{2}\int_{\delta }^{\infty }\mathrm{d}s\int \mathrm{d}^{4}x%
\sqrt{-g(x)}\left[ \widehat{H}_{1}H_{0}(s;x,x^{\prime };\xi )\right]
_{x^{\prime }=x}.  \label{gamma1}
\end{equation}
Thus, to compute the one-insertion one-loop diagrams it is sufficient to act
with $\widehat{H}_{1}$ on the unperturbed function $H_{0}$ and then take the
coincidence limit. The divergent part is given by the $\mathcal{O}(1/s)$
contributions to the square bracket in (\ref{gamma1}), namely 
\begin{equation}
\Gamma _{\mathrm{div}}^{(1)}=\frac{1}{4\varepsilon }\int \mathrm{d}^{4}x%
\sqrt{-g(x)}\left[ \widehat{H}_{1}H_{0}(s;x,x^{\prime };\xi )\right]
_{x^{\prime }=x}^{s^{-1}},  \label{namely}
\end{equation}
where the superscript $s^{-1}$ is to emphasize that only the coefficient of $%
1/s$ has to be kept, after inserting the Schwinger-DeWitt expansion for the
unperturbed function $H_{0}(s;x,x^{\prime };\xi )$.

To illustrate these facts in a simple example, consider a complex scalar
field $\varphi $ in curved space, described by the lagrangian 
\begin{equation}
\frac{\mathcal{L}}{\sqrt{-g}}=-\partial _{\mu }\overline{\varphi }g^{\mu \nu
}\partial _{\nu }\varphi -\xi R\overline{\varphi }\varphi +\overline{\varphi 
}\left( V+V^{\mu }D_{\mu }+V^{\mu \nu }D_{\mu }D_{\nu }+V^{\mu \nu \rho
}D_{\mu }D_{\nu }D_{\rho }+\cdots \right) \varphi ,  \label{lal}
\end{equation}
where all tensors $V^{\mu \nu \cdots }$ are symmetric.

The one-insertion divergent terms are then 
\begin{equation}
\Gamma _{\mathrm{div}}^{(1)}=\frac{1}{4\varepsilon }\int \mathrm{d}^{4}x%
\sqrt{-g}\left[ \left( V+V^{\mu }D_{\mu }+V^{\mu \nu }D_{\mu }D_{\nu
}+V^{\mu \nu \rho }D_{\mu }D_{\nu }D_{\rho }+\cdots \right)
H_{0}(s;x,x^{\prime };\xi )\right] _{x^{\prime }=x}^{s^{-1}}.  \label{oneins}
\end{equation}
The first two types of terms give immediately 
\[
\Gamma _{\mathrm{div}}^{(1)}=\frac{1}{64\pi ^{2}\varepsilon }\int \mathrm{d}%
^{4}x\sqrt{-g}\left( V\overline{A}_{1}+V^{\mu }\overline{A_{1;\mu }}\right) =%
\frac{(1-6\xi )}{24(4\pi )^{2}\varepsilon }\int \mathrm{d}^{4}x\sqrt{g}%
\left( VR+\frac{1}{2}V^{\mu }R_{;\mu }\right) . 
\]
The two-derivative term gives 
\begin{eqnarray*}
\Gamma _{\mathrm{div}}^{(1)} &=&\frac{1}{64\pi ^{2}\varepsilon }\int \mathrm{%
d}^{4}x\sqrt{-g}\left( V^{\mu \nu }\overline{A_{1;\mu \nu }}-\frac{1}{2}%
V_{\mu }^{\mu }\overline{A}_{2}\right) \\
&=&\frac{1}{960(4\pi )^{2}\varepsilon }\int \mathrm{d}^{4}x\sqrt{-g}\left[
4V^{\mu \nu }\left( \Box R_{\mu \nu }+(1-10\xi )RR_{\mu \nu }+(3-20\xi
)R_{;\mu \nu }-2R^{\rho \sigma }R_{\rho \mu \sigma \nu }\right) \right. \\
&&\left. \qquad \qquad \qquad \qquad \qquad -V_{\mu }^{\mu }\left( 2R_{\nu
\rho }R^{\nu \rho }+4(1-5\xi )\Box R+(60\xi ^{2}-20\xi +1)R^{2}\right)
\right] .
\end{eqnarray*}
The three-derivative term gives 
\begin{eqnarray}
\Gamma _{\mathrm{div}}^{(1)} &=&\frac{1}{64\pi ^{2}\varepsilon }\int \mathrm{%
d}^{4}x\sqrt{-g}\left( V^{\mu \nu \rho }\overline{A_{1;\mu \nu \rho }}-\frac{%
3}{2}V_{\mu }^{\mu \nu }\overline{A_{2;\nu }}\right)  \nonumber \\
&=&\frac{1}{64\pi ^{2}\varepsilon }\int \mathrm{d}^{4}x\sqrt{-g}\left[
\left( \frac{\xi }{8}-\frac{1}{40}\right) R_{;\mu \ \ \nu }^{\ \ \ \mu
}V_{\rho }^{\nu \rho }+\frac{1}{20}R^{\mu \rho }R_{\mu \nu \rho \sigma
}V_{;\alpha }^{\nu \sigma \alpha }+\frac{1}{80}R^{\mu \nu }R_{\mu \nu
}V_{\alpha ;\beta }^{\alpha \beta }\right.  \nonumber \\
&&+\left( \frac{1}{30}-\frac{\xi }{4}\right) R_{;\mu \nu \rho }V^{\mu \nu
\rho }+\frac{1}{40}R_{\mu \nu ;\alpha \ \rho }^{\ \ \ \ \ \alpha }V^{\mu \nu
\rho }+\left( \frac{\xi }{4}-\frac{1}{40}\right) R_{\mu \nu }RV_{;\rho
}^{\mu \nu \rho }  \nonumber \\
&&\left. -\left( \frac{1}{80}-\frac{\xi }{4}+\frac{3}{4}\xi ^{2}\right)
RR_{;\mu }V_{\nu }^{\nu \mu }\right] .  \label{m30}
\end{eqnarray}
The four-derivative term gives 
\[
\Gamma _{\mathrm{div}}^{(1)}=\frac{1}{64\pi ^{2}\varepsilon }\int \mathrm{d}%
^{4}x\sqrt{-g}\left( \overline{A_{1;\mu \nu \rho \sigma }}V^{\mu \nu \rho
\sigma }-3\overline{A_{2;\mu \nu }}V_{\rho }^{\mu \nu \rho }+\frac{3}{4}%
\overline{A_{3}}V_{\mu \nu }^{\mu \nu }\right) . 
\]
The coincidence limits $\overline{A_{3}}$, $\overline{A_{2;\mu \nu }}$ and $%
\overline{A_{1;\mu \nu \rho \sigma }}$ that are necessary to write this
expression explicitly have been worked out in \cite{decanini} and rederived
by ourselves.

The five- and six-derivative term gives 
\begin{eqnarray*}
\Gamma _{\mathrm{div}}^{(1)} &=&\frac{1}{64\pi ^{2}\varepsilon }\int \mathrm{%
d}^{4}x\sqrt{-g}\left( \overline{A_{1;\mu \nu \rho \sigma \alpha }}V^{\mu
\nu \rho \sigma \alpha }-5\overline{A_{2;\mu \nu \rho }}V_{\sigma }^{\mu \nu
\rho \sigma }+\frac{15}{4}\overline{A_{3;\rho }}V_{\mu \nu }^{\mu \nu \rho
}\right) , \\
\Gamma _{\mathrm{div}}^{(1)} &=&\frac{1}{64\pi ^{2}\varepsilon }\int \mathrm{%
d}^{4}x\sqrt{-g}\left( \overline{A_{1;\mu \nu \rho \sigma \alpha \beta }}%
V^{\mu \nu \rho \sigma \alpha \beta }-\frac{15}{2}\overline{A_{2;\mu \nu
\rho \sigma }}V_{\alpha }^{\mu \nu \rho \sigma \alpha }+\frac{45}{4}%
\overline{A_{3;\rho \sigma }}V_{\mu \nu }^{\mu \nu \rho \sigma }-\frac{15}{8}%
\overline{A_{4}}V_{\mu \nu \rho }^{\mu \nu \rho }\right) ,
\end{eqnarray*}
respectively. Apart from $\overline{A_{4}}$ , which has been computed in 
\cite{amster} and \cite{avramidi}, the unperturbed coefficients appearing in
these formulas have not been written in the literature.

\paragraph{\textit{A useful identity.}}

The formula for $\Gamma _{\mathrm{div}}^{(1)}$ can be simplified using the
identity 
\begin{equation}
\overline{\sigma _{;\lambda (\mu _{1}\cdots \mu _{n})}}=0,\quad \quad \qquad
\forall n>1,  \label{p1}
\end{equation}
where the parenthesis means complete symmetrization. Expressions such as $%
V^{\mu _{1}\cdots \mu _{n}}\overline{\sigma _{;\mu _{1}\cdots \mu _{n}}}$
and similar are thus identically zero. This property reduces the number of $%
\sigma $-derivatives that need to be computed to work out $\Gamma _{\mathrm{%
div}}^{(1)}$. Formula (\ref{p1}) can also be used to derive the coincidence
limits $\overline{\sigma _{;\mu _{1}\cdots \mu _{n}}}$ in a more efficient
way.

The proof of (\ref{p1}) can be done by induction. For $n=2$ the identity is
true, since $\overline{\sigma _{;\mu _{1}\mu _{2}\mu _{3}}}=0$. Assume that
it is true up to $n=\overline{n}>2$. Taking one derivative of the first
equation of (\ref{sigma}), we get 
\[
\sigma _{;\mu }=\sigma ^{;\lambda }\sigma _{;\lambda \mu }=\sigma ^{;\lambda
}\sigma _{;\mu \lambda }. 
\]
Now, take $\overline{n}+1$ derivatives of this equation and symmetrize
completely in those. We get 
\[
\sigma _{;\mu (\mu _{1}\cdots \mu _{\overline{n}+1})}=\sum_{k=0}^{\overline{n%
}+1}\left( 
\begin{array}{c}
\overline{n}+1 \\ 
k
\end{array}
\right)\sigma _{\quad (\mu _{1}\cdots \mu _{k}}^{;\lambda }\sigma _{;\mu
\lambda \ \mu _{k+1}\cdots \mu _{\overline{n}+1})}, 
\]
where $\lambda $ and $\mu $ are excluded from the symmetrization. Now, take
the coincidence limit of this expression and use the inductive hypothesis,
together with $\overline{\sigma _{;\mu }}=0$. The result simplifies to 
\[
\overline{\sigma _{;\mu (\mu _{1}\cdots \mu _{\overline{n}+1})}}=(\overline{n%
}+1)\overline{\sigma _{\quad (\mu _{1}}^{;\lambda }}\overline{\sigma _{;\mu
\lambda \ \mu _{2}\cdots \mu _{\overline{n}+1})}}+\overline{\sigma _{\quad
(\mu _{1}\cdots \mu _{\overline{n}+1})}^{;\lambda }}\overline{\sigma _{;\mu
\lambda }}. 
\]
Using $\overline{\sigma _{;\mu \nu }}=g_{\mu \nu }$ we arrive immediately at 
\[
\overline{\sigma _{;\mu (\mu _{1}\cdots \mu _{\overline{n}+1})}}=(\overline{n%
}+2)\overline{\sigma _{;\mu (\mu _{1}\cdots \mu _{\overline{n}+1})}}, 
\]
which proves the statement. We have checked (\ref{p1}) explicitly up to $n=7$
included, using the complete expressions of $\overline{\sigma _{;\mu
_{1}\cdots \mu _{n}}}$, $n\leq 8$, derived with a computer program.

\paragraph{\textit{The general formula.}}

Using (\ref{p1}) it is possible to work out the general formula 
\begin{equation}
\Gamma _{\mathrm{div}}^{(1)}=\frac{1}{64\pi ^{2}\varepsilon }\int \mathrm{d}%
^{4}x\sqrt{-g}\sum_{n=0}^{\infty }\sum_{k=0}^{[n/2]}\frac{(-1)^{k}n!}{%
4^{k}k!(n-2k)!}\overline{A_{k+1;\mu _{1}\cdots \mu _{n-2k}}}\mathrm{{tr}%
_{k}V^{\mu _{1}\cdots \mu _{n-2k}},}  \label{genfor}
\end{equation}
where $[n/2]$ is the integral part of $n/2$ and tr$_{k}V$ means that $k$
pairs of $V$-indices are traced. The formula is derived as follows. Consider
(\ref{oneins}) with the perturbation $V^{\mu _{1}\cdots \mu _{n}}D_{\mu
_{1}}\cdots D_{\mu _{n}}$: 
\[
\Gamma _{\mathrm{div}}^{(1)}=\frac{i}{4\varepsilon (4\pi )^{2}}\int \mathrm{d%
}^{4}x\sqrt{-g}\left\{ V^{\mu _{1}\cdots \mu _{n}}D_{\mu _{1}}\cdots D_{\mu
_{n}}\left[ \mathrm{\exp }\left( \frac{i\sigma (x,x^{\prime })}{2s}\right)
\sum_{n=0}^{\infty }(is)^{n-2}A_{n}(x,x^{\prime };\xi )\right] \right\}
_{x^{\prime }=x}^{s^{-1}}. 
\]
Since the derivatives are symmetrized, any time three or more of them act on 
$\sigma (x,x^{\prime })$ the contribution vanishes in the coincidence limit.
Moreover, since $\overline{\sigma }_{;\mu }$ vanishes, only two derivatives
can act on the same $\sigma $, all others having to act on the $A_{n}$'s.
Since $\overline{\sigma }_{\mu \nu }=g_{\mu \nu }$, two derivatives acting
on $\sigma $ trace a pair of $V$-indices. The remaining combinatorics are
then straightforward and give 
\[
\Gamma _{\mathrm{div}}^{(1)}=\frac{i}{64\pi ^{2}\varepsilon }\int \mathrm{d}%
^{4}x\sqrt{-g}\left\{ \sum_{k=0}^{[n/2]}\frac{(-1)^{k}n!}{2^{2k}k!(n-2k)!}%
\sum_{m=0}^{\infty }(is)^{m-2-k}\overline{A_{m;\mu _{1}\cdots \mu _{n-2k}}}%
\mathrm{{tr}_{k}V^{\mu _{1}\cdots \mu _{n-2k}}}\right\} ^{s^{-1}}, 
\]
which proves (\ref{oneins}).

In renormalization theory a further simplification applies.\ Indeed, it is
not necessary to include in (\ref{lal}) independent terms proportional to $%
\Box \varphi $, because they can be converted into terms of other types by
means of $\varphi $-field redefinitions. Up to terms proportional to the $%
\varphi $-field equations, $\Box \varphi $ can be replaced with 
\[
\xi R\varphi -(V+V^{\mu }D_{\mu }+V^{\mu \nu }D_{\mu }D_{\nu }+V^{\mu \nu
\rho }D_{\mu }D_{\nu }D_{\rho }+\cdots )\varphi . 
\]
Thus, a repeated use of field redefinitions can eliminate all boxes acting
on $\varphi $. Moreover, any couple of contracted $V$-indices can be moved
to a box acting on $\varphi $ commuting the covariant derivatives, up to $V$%
-terms with fewer indices. Thus, it is sufficient to take symmetric,
traceless $V$'s. The final result is then just 
\begin{equation}
\Gamma _{\mathrm{div}}^{(1)}=\frac{1}{64\pi ^{2}\varepsilon }\int \mathrm{d}%
^{4}x\sqrt{-g}\sum_{n=0}^{\infty }\overline{A_{1;\mu _{1}\cdots \mu _{n}}}%
\mathrm{{\ }V^{\mu _{1}\cdots \mu _{n}}.}  \label{notrace}
\end{equation}

\section{Calculation of total derivatives and multiple insertions}

\setcounter{equation}{0}

In a variety of computations, for example the short-distance expansion of
Green functions, total derivatives have to be kept. Moreover, with multiple $%
\widehat{H}_{1}$-insertions, even neglecting total derivatives, the
calculation simplifies much less than with single $\widehat{H}_{1}$%
-insertions. In this section we describe how the calculation based on the
CBH approach proceeds in the general case and report a number of sample and
new calculations of the perturbed coefficient $\overline{B_{2}}$. First we
focus on the total-derivative corrections to the single $H_{1}$-insertion
results computed in the previous section. Later we classify the structure of
contributions in the general case.

The scalar-potential, one-derivative and two-derivative results (\ref
{scalarpot}), (\ref{onederpot}) and (\ref{m2}) are known. They can be
derived in a variety of conventional ways. We rederive them with our
techniques to illustrate the CBH approach. They are also useful to introduce
the more difficult derivation of the three-derivative new result (\ref{m3}).

Write 
\[
H(s;x,x^{\prime };\xi )=H_{0}(s;x,x^{\prime };\xi )+H_{1}(s;x,x^{\prime
};\xi )+\mathcal{O}(H_{1}^{2}). 
\]
From the CBH\ formula (\ref{CBH}) we get 
\begin{equation}
H_{1}(s;x,x^{\prime };\xi )=\sum_{n=0}^{\infty }\frac{(is)^{n+1}}{(n+1)!}(%
\mathrm{ad}\widehat{H}_{0})^{n}\widehat{H}_{1}H_{0}(s;x,x^{\prime };\xi ).
\label{s2}
\end{equation}
Suppose that the interaction $\widehat{H}_{1}$ contains at most $m$
derivatives and that we are interested in the calculation of the coincidence
limit of a given perturbed coefficient, say $B_{k}$. Since $\widehat{H}_{0}$
contains at most two derivatives and each commutator $\mathrm{ad}\widehat{H}%
_{0}$ raises the number of derivatives by one unit, $(\mathrm{ad}\widehat{H}%
_{0})^{n}\widehat{H}_{1}$ contains at most $m+n$ derivatives acting on the
function $H_{0}(s;x,x^{\prime })$. When derivatives act on the exponential
prefactor $F\equiv \mathrm{exp}\left( \sigma (x,x^{\prime })/(2is)\right) $,
they lower the $s$-power.

Although each derivative acting on $F$ lowers the $s$-power by one unit, to
give a non-trivial contribution in the coincidence limit derivatives have to
act on $F$ at least in pairs, because $\overline{\sigma _{;\mu }}=0$. Thus,
in the coincidence limit, the $s$-power can be lowered by at most $[(m+n)/2]$
units. The lowest $s$-power that multiplies the unperturbed coefficient $%
A_{k^{\prime }}$ is 
\begin{equation}
n+1-\left[ \frac{m+n}{2}\right] +k^{\prime }-2.  \label{isto}
\end{equation}
Here the factor $s^{n+1}$ comes from (\ref{s2}), while $s^{k^{\prime }-2}$
multiplies $A_{k^{\prime }}$ inside $H_{0}(s;x,x^{\prime })$. Now, in the
Schwinger-DeWitt expansion (\ref{sdw}) of the function $H(s;x,x^{\prime
};\xi )$, the coefficient $B_{k}$ is multiplied by $k-2$ powers of $s$.
Equating this number (\ref{isto}), we see that, for the purpose of computing
the single-insertion perturbations to $B_{k}$, the sum in (\ref{s2}) becomes
finite. It stops at the $\overline{n}$ such that 
\begin{equation}
\overline{n}+1-\left[ \frac{m+\overline{n}}{2}\right] =k.  \label{diof}
\end{equation}
With multiple insertions, say $j$, the sum of (\ref{sum}) is raised to the
power $j$. Call $n$ the total power of $\mathrm{ad}\widehat{H}_{0}$
contributing from the $\widetilde{H}_{1}(\zeta )$s and $m_{j}$ the total
number of derivatives carried by the $j$ insertions. Then equation (\ref
{diof}) is generalized to 
\[
j+\overline{n}-\left[ \frac{m_{j}+\overline{n}}{2}\right] =k, 
\]
The list of contributions stops when the total power of $\mathrm{ad}\widehat{%
H}_{0}$ reaches the value $\overline{n}$.

This counting proves that the CBH method is consistent with the perturbative
expansion, and the coincidence limit of each perturbed Schwinger-DeWitt
coefficient can be calculated algorithmically. However, the calculation can
become lengthy quite soon, even for computer programs. We now compute the
single-insertion perturbations to $\overline{B_{2}}$ for the cases
considered in the previous section and classify the contributions of
multiple insertions in more detail.

\paragraph{\textit{Scalar-potential perturbation.}}

The simplest perturbation is the scalar potential $V(x)$. In formula (\ref
{s2}) the term with $n=0$ has one power of $s$, so it gives a contributions
proportional to $\overline{A_{1}}$. The term with $n=1$, contains two powers
of $s$ and at most one derivative acting on $H_{0}(s;x,x^{\prime })$, so its
contributions are proportional to $\overline{A_{0;\mu }}$, which vanishes,
and $\overline{A_{0}}=1$. The term with $n=2$ contains three powers of $s$
and at most two derivatives acting on $H_{0}$, which lower the $s$-power by
at most one unit when they act on the prefactor $F$. This contribution is
again proportional to $\overline{A_{0}}$, and $\Box V$. The terms with $%
n\geq 3$ do not contribute, because they contain too many powers of $s$ and
too few derivatives to lower them. Working out the commutators we get 
\begin{equation}
\Delta \overline{B_{2}}=\left( \frac{1}{6}-\xi \right) RV+\frac{1}{6}\Box V.
\label{scalarpot}
\end{equation}

\paragraph{\textit{One-derivative perturbation.}}

Now, consider the perturbation $V^{\mu }D_{\mu }$. The first term of (\ref
{s2}) gives a contribution proportional to $\overline{A_{1;\mu }}$. The
second term contains two powers of $s$ and at most two derivatives acting on 
$H_{0}$, thus it gives contributions proportional to $\overline{A_{0;2}}$
and $\overline{A_{1}}$, where $\overline{A_{0;2}}$ denotes any object with
less than three derivatives, namely $\overline{A_{0;\mu \nu }}$, $\overline{%
A_{0;\mu }}$ and $\overline{A_{0}}$. The third term has at most three
derivatives on $H_{0}$. Two of them are used to lower one $s$ power, so the
contribution is proportional to $\overline{A_{0;1}}$. The third term has
four derivatives that have to be used to lower the $s$-power by two units,
giving a contribution proportional to $\overline{A_{0}}$. The terms of (\ref
{s2}) with $n\geq 4$ do not contribute, because they have too many $s$'s and
too few derivatives acting on $H_{0}$. The result is 
\begin{equation}
\Delta \overline{B_{2}}=-\frac{1-6\xi }{12}RV_{;\mu }^{\mu }-\frac{1}{12}%
\Box V_{;\mu }^{\mu }.  \label{onederpot}
\end{equation}

We know that the terms with $n\geq 1$ are total derivatives. Using the
differential approach it is not easy to recognize the presence of such total
derivatives, which are often very involved. The simplest of them is 
\[
\langle x\mid \mathrm{e}^{i\widetilde{H}_{0}s}(\mathrm{ad}\widetilde{H}_{0})%
\widetilde{H}_{1}\mid x\rangle .
\]
Let us inspect it more closely, to see what kind of relations it generates.
First observe that we can always move the covariant derivatives away from $%
V^{\mu }$, eventually adding other total derivatives. When we do this, we
get a relation of the form 
\[
V^{\mu }J_{\mu }=\mathrm{total\,\,derivative,}
\]
for some current $J^{\mu }$. Next, integrating this relation over spacetime
and using the arbitrariness of $V^{\mu }$, we obtain the identity $J_{\mu }=0
$. Finally, substituting the $\sigma $-coincidence limits, we obtain a
relation for the $A_{k}$-coincidence limits. The result is 
\begin{equation}
0=D_{\mu }\overline{A_{1}}+\xi \overline{A_{0}}R_{;\mu }+\Box \overline{%
A_{0;\mu }}-2D^{\nu }\overline{A_{0;\mu \nu }}+\overline{A_{0;\nu }}R_{\mu
}^{\nu }.  \label{idi}
\end{equation}
Notice that some derivatives are taken before the coincidence limits, others
are taken after the coincidence limits. The values of $\overline{A_{0;\mu
\nu }}$, $\overline{A_{0;\mu }}$, $\overline{A_{0}}$ and $\overline{A_{1}}$
are reported in the appendix and indeed satisfy (\ref{idi}). More
complicated identities are generated by the other terms of (\ref{s2}).

\paragraph{\textit{Two-derivative perturbation.}}

Let us consider the perturbation $V^{\mu \nu }D_{\mu }D_{\nu }$. The term
with $n=0$ in (\ref{s2}) contains two derivatives, that can either act on
the exponential prefactor of $H_{0}$, lowering the $s$-power by one unit, or
on the unperturbed Schwinger-DeWitt coefficients contained in the expansion
of $H_{0}$. The resulting contribution is a linear combination of $\overline{%
A_{2}}$ and $\overline{A_{1;2}}$. The term with $n=1$ contains at most three
derivatives acting on $H_{0}$, two of which can act on the exponential
prefactor. The result is a sum of $\overline{A_{1;1}}$ and $\overline{A_{0;3}%
}$. The third term of (\ref{s2}) contains at most four derivatives on $H_{0}$%
. Four or two of them can lower the $s$-power by two units or one,
respectively. The contributions of this term are proportional to $\overline{%
A_{1}}$ and $\overline{A_{0;2}}$. Similarly, the terms with $n=3$ and $n=4$
give contributions proportional $\overline{A_{0;1}}$ and $\overline{A_{0}}$,
respectively.

The final result is given by 
\begin{eqnarray}
\Delta \overline{B_{2}} &=&\frac{1-6\xi }{18}RV_{;\mu \nu }^{\mu \nu }+\frac{%
1-5\xi }{30}D_{\nu }(R_{;\mu }V^{\mu \nu })+\frac{1}{36}R_{\mu \nu }\Box
V^{\mu \nu }+\frac{1}{90}R_{\mu }^{\nu }V_{;\nu \rho }^{\mu \rho }+\frac{1}{%
15}R_{\mu \nu ;\rho }V^{\mu \nu ;\rho }  \nonumber \\
&&+\frac{1}{30}\Box R_{\mu \nu }V^{\mu \nu }+\frac{1-10\xi }{60}R_{\mu \nu
}RV^{\mu \nu }+\frac{1}{90}R_{\rho \sigma }R_{\mu }^{\sigma }V^{\rho \mu }-%
\frac{1}{45}R_{\mu \nu \rho \sigma }V^{\mu \rho ;\nu \sigma }+\frac{1}{20}%
\Box V_{;\mu \nu }^{\mu \nu }  \nonumber \\
&&-\frac{1}{45}R_{\mu \nu }R^{\mu \rho \nu \sigma }V_{\rho \sigma }-\frac{%
1-5\xi }{60}R_{;\mu }\mathrm{{tr}V^{;\mu }-\frac{1}{180}R_{\mu \nu }{tr}%
V^{;\mu \nu }-\frac{1-6\xi }{72}R\Box {tr}V-\frac{1}{120}\Box ^{2}{tr}V} 
\nonumber \\
&&-\mathrm{{tr}V\left( \frac{1-20\xi +60\xi ^{2}}{240}R^{2}+\frac{1}{120}%
R_{\mu \nu }R^{\mu \nu }+\frac{1-5\xi }{60}\Box R\right) .}  \label{m2}
\end{eqnarray}

\paragraph{$m$-\textit{derivative perturbation.}}

In the general case, namely a perturbation $V^{\mu _{1}\cdots \mu
_{m}}D_{\mu _{1}}\cdots D_{\mu _{m}}$, the first contribution of (\ref{s2})
gives the list of terms written in (\ref{genfor}). Each commutator with $%
\widehat{H}_{0}$ in (\ref{s2}) raises the $s$-power by one unit and the
number of derivatives by one unit. If the new derivative does not act on the
exponential prefactor of $H_{0}$, we have 
\begin{equation}
\overline{A_{k;j}}\rightarrow \overline{A_{k-1;j+1}}.  \label{op1}
\end{equation}
If the derivative acts on the exponential prefactor, then it must absorb a
second derivative, to give a non-trivial contribution. In this case, both
the $s$-power and the number of derivatives are lowered by one unit: 
\begin{equation}
\overline{A_{k;j}}\rightarrow \overline{A_{k;j-1}}.  \label{op2}
\end{equation}
Combining the two operations the contributions fit into the following
scheme: 
\[
\begin{tabular}{|c|ccccc|}
\hline
$n=0$ & $\overline{A_{[m/2]+1;\sigma (m)}}$ & $\overline{A_{[m/2];\sigma
(m)+2}}$ & $\cdots $ & $\overline{A_{1;m}}$ &  \\ 
$n=1$ & $\overline{A_{[m/2]+1;\sigma (m)-1}}$ & $\overline{A_{[m/2];\sigma
(m)+1}}$ & $\cdots $ & $\overline{A_{1;m-1}}$ & $\overline{A_{0;m+1}}$ \\ 
$n=2$ &  & $\overline{A_{[m/2];\sigma (m)}}$ & $\cdots $ & $\cdots $ & $%
\cdots $ \\ 
$n=3$ &  & $\overline{A_{[m/2];\sigma (m)-1}}$ & $\cdots $ & $\cdots $ & $%
\cdots $ \\ 
$\cdots $ &  &  & $\cdots $ & $\cdots $ & $\cdots $ \\ 
$n=m$ &  &  &  & $\overline{A_{1;0}}$ & $\overline{A_{0;2}}$ \\ 
$n=m+1$ &  &  &  &  & $\overline{A_{0;1}}$ \\ 
$n=m+2$ &  &  &  &  & $\overline{A_{0;0}}$ \\ \hline
\end{tabular}
\]
Here $\sigma (m)=0$ if $m$ is even, $\sigma (m)=1$ if $m$ is odd. Observe
that the coefficient $\overline{A_{0;m+2}}$, the most involved of all, does
not contribute.

For example, for $m=3$ we have contributions 
\[
\begin{tabular}{|c|ccc|}
\hline
$n=0$ & $\overline{A_{2;1}}$ & $\overline{A_{1;3}}$ &  \\ 
$n=1$ & $\overline{A_{2}}$ & $\overline{A_{1;2}}$ & $\overline{A_{0;4}}$ \\ 
$n=2$ &  & $\overline{A_{1;1}}$ & $\overline{A_{0;3}}$ \\ 
$n=3$ &  & $\overline{A_{1}}$ & $\overline{A_{0;2}}$ \\ 
$n=4$ &  &  & $\overline{A_{0;1}}$ \\ 
$n=5$ &  &  & $\overline{A_{0}}$ \\ \hline
\end{tabular}
\]
Each of these coefficients are available in the literature and have been
recalculated independently by us. The $m=3$ perturbed coefficient reads: 
\begin{eqnarray}
&&\Delta \overline{B_{2}}=\frac{1}{40}\left( \frac{1}{2}\Box ^{2}+\frac{1}{2}%
R_{\mu \nu }R^{\mu \nu }+\frac{5}{6}R\Box +\frac{1}{4}R^{2}+\Box R+R_{;\mu
}D^{\mu }+\frac{1}{3}R_{\mu \nu }D^{\mu }D^{\nu }\right) V_{\rho ;\sigma
}^{\rho \sigma }  \nonumber \\
&&-\frac{1}{6}\left( \frac{1}{4}R+\frac{1}{5}\Box \right) V_{;\mu \nu \rho
}^{\mu \nu \rho }-\frac{1}{20}R_{;\mu }V_{;\nu \rho }^{\mu \nu \rho }-\frac{1%
}{60}R_{\mu }^{\nu }V_{;\nu \rho \sigma }^{\mu \rho \sigma }-\frac{1}{60}%
R_{\mu \nu }R^{\mu \rho \sigma \alpha }V_{\rho \sigma ;\alpha }^{\nu }-\frac{%
1}{24}R_{\mu \nu }\Box V_{;\alpha }^{\mu \nu \alpha }  \nonumber \\
&&-\frac{1}{10}\left( \frac{1}{2}\Box R_{\mu \nu }+\frac{1}{4}RR_{\mu \nu }+%
\frac{1}{2}R_{;\mu \nu }+R_{\mu \nu ;\rho }D^{\rho }-\frac{1}{3}R_{\mu \rho
\nu \sigma }D^{\rho }D^{\sigma }-\frac{1}{2}R^{\rho \sigma }R_{\mu \rho \nu
\sigma }\right) V_{;\alpha }^{\mu \nu \alpha }  \nonumber \\
&&+\frac{1}{30}\left( R_{\mu }^{\nu }R_{\nu \rho ;\sigma }+\frac{1}{2}R^{\nu
\alpha }R_{\nu \rho \alpha \sigma ;\mu }-\frac{1}{2}R_{\mu }^{\nu }R_{\rho
\sigma ;\nu }\right) V^{\mu \rho \sigma }+\frac{\xi }{4}RV_{;\mu \nu \rho
}^{\mu \nu \rho }+\frac{\xi }{4}R_{;\mu }V_{;\nu \rho }^{\mu \nu \rho } 
\nonumber \\
&&-\frac{\xi }{8}\left( R\Box +(1-3\xi )R^{2}+\Box R+R_{;\mu }D^{\mu
}\right) V_{\rho ;\sigma }^{\rho \sigma }+\frac{\xi }{4}\left( RR_{\mu \nu
}+R_{;\mu \nu }\right) V_{;\alpha }^{\mu \nu \alpha }.  \label{m3}
\end{eqnarray}
The unperturbed coefficients necessary for the $m=4$ result exist in the
literature \cite{decanini}. For $m>4$ the necessary coefficients can be
derived with computer programs, but an increasing amount of time is required.

\paragraph{\textit{Squared-Laplacian perturbation.}}

An interesting case is the perturbation $\lambda \Box ^{2}$, which is a
linear combination of the perturbations $V^{\mu _{1}\cdots \mu _{m}}D_{\mu
_{1}}\cdots D_{\mu _{m}}$ with $m=1,2,4$. The commutators are much simpler
in this case and the final result is

\[
\Delta \overline{B_{2}}=\lambda \xi \left[ \frac{1}{15}R_{;\mu }R^{;\mu }+%
\frac{1}{45}R^{\mu \nu }R_{;\mu \nu }+\frac{1}{30}\Box ^{2}R+\frac{1}{18}%
R\Box R+\xi \left( \frac{1}{6}-\xi \right) R^{3}\right] . \label{squaredlapl}
\]

\paragraph{\textit{Multiple insertions.}}

So far, we have classified the single $\widehat{H}_{1}$-insertions, but the
analysis can be generalized to more insertions. We do it for the calculation
of a generic perturbed coefficient $\overline{B_{k}}$. Each
multiple-insertion contribution is made by a certain number of $\widehat{H}%
_{1}$'s and a certain number of $($ad$\widehat{H}_{0})$'s acting on them.
Call $n$ the ``level'' of the contribution, namely the total number of $%
\widehat{H}_{0}$-commutators. Denote the total number of $\widehat{H}_{1}$%
-insertions with $r$. In each insertion, pick a perturbation $V^{\mu
_{1}\cdots \mu _{m}}D_{\mu _{1}}\cdots D_{\mu _{m}}$, not necessarily with
the same number $m$ of derivatives. Call $d$ the total number of derivatives
carried by such perturbations. Since $\widehat{H}_{0}$ has two derivatives
at most, the total number of derivatives acting on $H_{0}$ is at most equal
to $d+n$. Acting on the exponential prefactor of $H_{0}$, such derivatives
can lower the $s$-power by at most $[(d+n)/2]$ units. We get non-vanishing
contributions when $n_{\max }=d+2(k-r)\geq 0$. They are proportional to the
unperturbed coefficients 
\begin{equation}
\overline{A_{k-r-n+[(d+n)/2];\sigma (d+n)}},\qquad \overline{%
A_{k-r-n-1+[(d+n)/2];\sigma (d+n)+2}}\qquad \cdots \qquad \overline{%
A_{k-r-n;d+n}}\ ,  \label{classa}
\end{equation}
where $n=0,1,\cdots ,n_{\max }$. In (\ref{classa}) $\overline{A_{p;q}}$ is
meant to vanish whenever $p<0$.

The classification applies to any unperturbed two-derivative operator $%
\widehat{H}_{0}$, in particular the spin-2 operator defined by gravity
expanded around an arbitrary background. Finally, it can be easily
generalized to operators $\widehat{H}_{0}$ with a different maximal number
of derivatives, to include fermions.

\paragraph{\textit{Non-minimal terms.}}

Even if they are not multiplied by ``small'' parameters, non-minimal terms
can be treated as perturbations, included in the $n=0$ term of $\widehat{H}%
_{1}$ in (\ref{h1}). Indeed, each coefficient of the Schwinger-DeWitt
expansion receives contributions from a finite number of non-minimal
insertions and a finite number of commutators with $\widehat{H}_{0}$.
Moreover, because non-minimal terms do not contain derivatives, their
contributions are relatively easy to compute. For example to compute $%
\overline{B_{2}}$ for 
\[
\widehat{H}=\Box +V, 
\]
we can apply (\ref{classa}) with $k=2$ and $d=0$. We obtain non-vanishing
contributions for $n=0,1,2$, $r=1,2$, namely a linear combination of $V^{2}%
\overline{A_{0}}$, $V\overline{A_{0}}$, $V\overline{A_{1}}$ and $V\overline{%
A_{0;1}}$.

\section{The case of gravity}

\setcounter{equation}{0}

Expanding (\ref{corra}) around a background metric and choosing the harmonic
gauge, the unperturbed spin-2 operator has the form 
\[
\widehat{H}_{0\mu \nu }^{\quad \rho ^{\prime }\sigma ^{\prime }}=\Box \left(
P_{2\hspace{0.01in}\mu \nu }^{~~~\rho ^{\prime }\sigma ^{\prime }}-P_{0%
\hspace{0.01in}\mu \nu }^{~~~\rho ^{\prime }\sigma ^{\prime }}\right) +%
\mathrm{nonminimal}\,\,\mathrm{terms,}
\]
where 
\[
P_{2\hspace{0.01in}\mu \nu }^{~~~\rho ^{\prime }\sigma ^{\prime }}=\frac{1}{2%
}\left( \delta _{\mu }^{\rho ^{\prime }}\delta _{\nu }^{\sigma ^{\prime
}}+\delta _{\mu }^{\rho ^{\prime }}\delta _{\nu }^{\sigma ^{\prime }}-\frac{1%
}{2}g_{\mu \nu }g^{\rho ^{\prime }\sigma ^{\prime }}\right) ,\qquad P_{0%
\hspace{0.01in}\mu \nu }^{~~~\rho ^{\prime }\sigma ^{\prime }}=\frac{1}{4}%
g_{\mu \nu }g^{\rho ^{\prime }\sigma ^{\prime }},
\]
are the projectors on the traceless and trace components, respectively.
Define the bitensor $H_{0\mu \nu }^{\quad \rho ^{\prime }\sigma ^{\prime
}}(s;x,x^{\prime })$ as the solution of 
\begin{equation}
i\frac{\partial }{\partial s}H_{0\mu \nu }^{\quad \rho ^{\prime }\sigma
^{\prime }}(s;x,x^{\prime })+\widehat{H}_{0\mu \nu }^{\quad \alpha \beta
}H_{0\alpha \beta }^{\quad \rho ^{\prime }\sigma ^{\prime }}(s;x,x^{\prime
})=0,  \label{eq1}
\end{equation}
with the boundary condition 
\begin{equation}
H_{0\mu \nu }^{\quad \rho ^{\prime }\sigma ^{\prime }}(0;x,x^{\prime })=%
\frac{\delta _{\mu }^{\rho ^{\prime }}\delta _{\nu }^{\sigma ^{\prime
}}+\delta _{\mu }^{\rho ^{\prime }}\delta _{\nu }^{\sigma ^{\prime }}}{2%
\sqrt{-g(x)}}\delta ^{(4)}(x-x^{\prime }).  \label{eq2}
\end{equation}
Write the Schwinger-DeWitt expansion of $H_{0\mu \nu }^{\quad \rho ^{\prime
}\sigma ^{\prime }}(s;x,x^{\prime })$ as 
\[
H_{0\mu \nu }^{\quad \rho ^{\prime }\sigma ^{\prime }}(s;x,x^{\prime })=-%
\frac{i}{(4\pi )^{2}s^{2}}\mathrm{\exp }\left( \frac{i\sigma (x,x^{\prime })%
}{2s}\right) \sum_{n=0}^{\infty }(is)^{n}A_{n\hspace{0.01in}\mu \nu }^{\quad
\rho ^{\prime }\sigma ^{\prime }}(x,x^{\prime }).
\]
The most general higher-derivative perturbation can be written as 
\begin{equation}
\widehat{H}_{1\mu \nu }^{\quad \rho ^{\prime }\sigma ^{\prime
}}=\sum_{n=0}^{\infty }V_{\mu \nu }^{\quad \rho ^{\prime }\sigma ^{\prime
}|\mu _{1}\cdots \mu _{n}}D_{\mu _{1}}\cdots D_{\mu _{n}},  \label{perta2}
\end{equation}
where $V_{\mu \nu }^{\quad \rho ^{\prime }\sigma ^{\prime }|\mu _{1}\cdots
\mu _{n}}$ are completely symmetric tensors in the indices $\mu _{1}\cdots
\mu _{n}$, while the other indices satisfy obvious symmetry properties. The
CBH\ approach described in this paper can be applied with virtually no
change. For example, in the case of a single insertion formula (\ref{gamma1}%
) generalizes to 
\[
\Gamma ^{(1)}=\frac{1}{2}\int_{\delta }^{\infty }\mathrm{d}s\int \mathrm{d}%
^{4}x\sqrt{-g(x)}\left[ \widehat{H}_{1\mu ^{\prime }\nu ^{\prime }}^{\quad
\rho \sigma }H_{0\rho \sigma }^{\quad \mu ^{\prime }\nu ^{\prime
}}(s;x,x^{\prime })\right] _{x^{\prime }=x}
\]
and (\ref{genfor}) becomes 
\[
\Gamma _{\mathrm{div}}^{(1)}=\frac{1}{64\pi ^{2}\varepsilon }\int \mathrm{d}%
^{4}x\sqrt{-g}\sum_{n=0}^{\infty }\sum_{k=0}^{[n/2]}\frac{(-1)^{k}n!}{%
4^{k}k!(n-2k)!}\overline{A_{k+1\hspace{0.01in}\rho \sigma ;\mu _{1}\cdots
\mu _{n-2k}}^{\qquad \mu ^{\prime }\nu ^{\prime }}}\mathrm{{tr}_{k}V_{\mu
^{\prime }\nu ^{\prime }}^{\quad \rho \sigma |\mu _{1}\cdots \mu _{n-2k}}.}
\]
For renormalization purposes $V_{\mu ^{\prime }\nu ^{\prime }}^{\quad \rho
\sigma |\mu _{1}\cdots \mu _{n}}$ can be taken to be traceless in $\mu
_{1}\cdots \mu _{n}$, which amounts to exclude terms proportional to the
field equations in (\ref{perta2}). Then formula (\ref{notrace}) becomes 
\[
\Gamma ^{(1)}=\frac{1}{64\pi ^{2}\varepsilon }\int \mathrm{d}^{4}x\sqrt{-g}%
\sum_{n=0}^{\infty }\overline{A_{1\hspace{0.01in}\rho \sigma ;\mu _{1}\cdots
\mu _{n}}^{\qquad \mu ^{\prime }\nu ^{\prime }}}\ V_{\mu ^{\prime }\nu
^{\prime }}^{\quad \rho \sigma |\mu _{1}\cdots \mu _{n}}.
\]

\section{Conclusions}

\setcounter{equation}{0}

In this paper we have studied improved Schwinger-DeWitt techniques for
higher-derivative perturbations of operator determinants and Green
functions, to calculate counterterms and short-distance expansions of
Feynman diagrams. In the perturbative regime the differential approach
presents some difficulties, but it can be efficiently superseded by a
systematic use of the CBH formula. We have classified the contributions that
arise in this framework and outlined a number of simplification techniques.
In some cases the calculational effort reduces considerably, in particular
when total derivatives can be neglected. The procedure is very general and
applies also to quantum gravity treated with the background field method.

Certain identities, such as (\ref{p1}), are new results, to our knowledge.
They have been used to derive the closed formulas (\ref{genfor}) and (\ref
{notrace}) that relate the most general single-insertion perturbed
Schwinger-DeWitt coefficients to the unperturbed ones, up to total
derivatives. Another new result is the three-derivative one-loop perturbed
coefficient (\ref{m3}). When total derivatives are included and/or multiple
insertions are considered, the list of contributing unperturbed coefficients
becomes considerably long. Nevertheless, we point out the simplicity of
formulas (\ref{m2}) and (\ref{m3}), compared with the involved intermediate
expressions that lead to them. In particular, the inclusion of total
derivatives in (\ref{m3}) does not make the result much more complicated
than (\ref{m30}), because several terms of (\ref{m30}) are canceled by the
total-derivative contributions. These facts suggest that there should exist
more powerful and systematic simplification methods than the ones uncovered
here. Hopefully the techniques of this paper can be extended and combined
with the background field method to study two-loop and higher-order
radiative corrections.

\vskip 25truept \noindent {\Large \textbf{Acknowledgments}}

\vskip 15truept \noindent

One of us (D.A.) would like to thank P. Menotti for useful discussions.

\vskip 25truept \noindent {\Large \textbf{A\ \ Appendix: conventions}}

\vskip 15truept

\renewcommand{\theequation}{A.\arabic{equation}} \setcounter{equation}{0}

\noindent The conventions used in this paper are those dubbed ``SecondUp''
(i.e. the default ones) in the package Ricci \cite{ricci} with metric
signature $(+,+,+,-)$. Precisely, if $V_{\mu }$ is a vector, 
\[
V_{\mu ;\nu \rho }-V_{\mu ;\rho \nu }=R_{\mu \sigma \nu \rho }V^{\sigma
},\qquad R_{\mu \nu }=R_{\ \mu \nu \rho }^{\rho },\qquad R=R_{\mu }^{\mu }. 
\]
We have performed our computations with two independent methods. The first
method used a Mathematica package written by one of us (D.A.), the second
method used the Ricci package.

The first few Schwinger-DeWitt coefficients in the coincidence limit are 
\[
\overline{A_{0}}=1,\quad \overline{A_{0;\mu }}=0,\quad \overline{A_{0;\mu
\nu }}=\frac{1}{6}R_{\mu \nu },\quad \overline{A_{1}}=\frac{1-6\xi }{6}%
R,\quad \overline{A_{1;\mu }}=\frac{1-6\xi }{12}R_{;\mu }. 
\]

\end{document}